\documentclass[11pt]{amsart}
\usepackage{amsmath, amssymb}
\usepackage{upgreek}
\usepackage{xcolor, url}
\usepackage[all]{xy}
\usepackage{graphicx}
\usepackage{bm}
\usepackage{enumerate}
\theoremstyle{plain}

\newcommand{\td}{\text{d}}
\theoremstyle{definition}

\numberwithin{equation}{section}

\newcommand{\tr}{\operatorname{Tr}}

\usepackage{subfig, tikz}
\usetikzlibrary{decorations.pathmorphing}
\usepackage{graphics}

\usepackage{hyperref} 
\hypersetup{colorlinks=true, linkcolor=blue, citecolor = blue, urlcolor=blue, filecolor=magenta}

\allowdisplaybreaks

\begin{document}
\title[On the Chen-Teo family of stationary asymptotically locally Minkowskian black holes]{On the Chen-Teo family of stationary asymptotically locally Minkowskian black holes}
\author{Federico Elizondo Lopez}
\address{Department of Physics and Physical Oceanography\\
		Memorial University of Newfoundland\\
		St John's, NL Canada}
	\email{aelizondolop@mun.ca}
\author{Hari K. Kunduri}
\address{Department of Mathematics and Statistics and Department of Physics and Astronomy\\
		McMaster University\\
		Hamilton, ON Canada}
	\email{kundurih@mcmaster.ca}
\author{Hakim Temacini}
\address{Department of Mathematics and Statistics\\
		McMaster University\\
		Hamilton, ON Canada}
	\email{temacinh@mcmaster.ca}

\thanks{H. K. Kunduri acknowledges the support of the  NSERC Grant  RGPIN-2018-04887. H. Temacini is supported by an NSERC 2025 summer research fellowship. }

\begin{abstract}
Chen and Teo have constructed a two-parameter family of five dimensional, stationary vacuum black hole solutions whose spatial hypersurfaces are asymptotically locally Euclidean with boundary at infinity $L(2,1)$. Spatial cross sections of the event horizon have topology $S^3$ equipped with inhomogeneous metrics.  When the mass is zero, the solution reduces to the trivial product of time with the Eguchi-Hanson gravitational instanton. We show that the spacetime metric can be smoothly extended through an event horizon and that the exterior region is stably causal. We also investigate their geometric and physical properties. In particular, we show that the Smarr relation and first law of black hole mechanics hold and compute the renormalized gravitational action. 
\end{abstract}

\maketitle
\section{Introduction}
In four dimensional general relativity, isolated systems in equilibrium are described by stationary spacetimes which asymptotically approach Minkowski spacetime in a suitable sense \cite{Chrusciel:2012jk}. There is a natural time function which approaches the standard Minkowski time coordinate, and its level sets will be spacelike hypsersurfaces that are asymptotically Euclidean (they have an asymptotic end diffeomorphic to $\mathbb{R}^3 \setminus \textrm{Ball}$). The classic black hole uniqueness theorems for stationary, asymptotically flat black holes assert that the Kerr family exhausts the set of such (analytic) vacuum solutions (for a precise statement, see \cite{Chrusciel:2008js}). These uniqueness results extend to five dimensions under the assumption of an additional rotational symmetry~\cite{Hollands:2007aj, Hollands:2008fm} (in the static case, no such assumption is required \cite{Gibbons:2002av}). 

It is natural to consider the existence of black hole solutions which asymptotically approach a different vacuum `background' of the form \begin{equation} \label{soliton} \mathbf{g} = -\td t^2 + g_b,\end{equation}  where $(M,g_b)$ is a complete, non-compact Riemannian manifold. In spacetime dimension $d=4$, Ricci flatness implies $g_b$ is Ricci flat, and hence $(M,g_b)$ is locally isometric to Euclidean $\mathbb{R}^3$. However, for $d=5$, Ricci flatness of $g_b$ allows for a richer set of possibilities; such spaces are known as \emph{gravitational instantons} \cite{Gibbons:1979xm}. Historically, such geometries were studied in the context of Euclidean quantum gravity (see \cite{Witten:2024upt} for a recent review), and a host of examples can be constructed by studying analytic continuations of Lorentizian black holes metrics such as Schwarzschild and Kerr. Other examples can be constructed directly by focusing on solutions with special geometry (e.g., the self-dual Taub-NUT and Eguchi-Hanson instantons, which are particular examples of a general set of hyperK\"ahler Gibbons-Hawking solutions \cite{Gibbons:1978tef}). There has been a recent renewed interest in gravitational instantons from the geometric perspective, following the explicit construction by Chen-Teo of a new family of gravitational instanton solutions on $\mathbb{CP}^2 \setminus S^1$ \cite{Chen:2011tc}. This has led to powerful classification results in the toric Hermitian setting~\cite{Biquard:2021gwj} and under the assumption of $S^1$-invariance \cite{Aksteiner:2023djq}, and existence and uniqueness results in the toric setting \cite{Kunduri:2021xiv}.

Constructing explicit black hole solutions which are asymptotic to Lorentzian metrics of the form \eqref{soliton} for a given gravitational instanton $g_b$ is a difficult task (to avoid confusion we will refer to the Lorentzian spacetimes $(\mathbb{R} \times M_b, \mathbf{g})$ associated to a given instanton $(M_b, g_b)$ as their corresponding \emph{gravitational solitons}).  Examples can be produced by restricting to solutions with special geometry. In particular, supersymmetric  (i.e. solutions admitting Killing spinors) spherical black holes and black rings asymptotic to the Taub-NUT \cite{Elvang:2005sa} and Eguchi-Hanson solitons \cite{Tomizawa:2007he,Tomizawa:2008tj} can be constructed\footnote{Note that Kaluza Klein spacetime with $M_b = S^1 \times \mathbb{R}^3$ with $g_b$ its canonical flat metric can also be considered; for a classifcation of supersymmetric black holes in this setting see \cite{Katona:2023uaj}.}.
These are black hole solutions of five dimensional supergravity and are necessarily extreme (their event horizon is degenerate) and carry electric charge. 

To construct stationary \emph{vacuum} black holes in five dimensions, restricting to biaxisymmetric solutions (those admitting a torus $\mathbb{T}^2 = U(1) \times U(1)$ isometry subgroup) has proved extremely useful. In this setting, the Einstein equations can be reformulated as a harmonic map from a two-dimensional orbit space $B:=M / (\mathbb{R} \times \mathbb{T}^2)$ into a non-positively curved target space. In particular $B$ is a two dimensional manifold with boundary diffeomorphic to the upper half plane $\mathbb{R}^2_+ = \{(\rho,z) | \rho \geq  0, z \in \mathbb{R} \}$ and the spacetime metric can be expressed in the interior as \cite{Hollands:2007aj}
\begin{equation}
    \mathbf{g} = e^{2\nu} (\td \phi^2 + \td z^2) + G_{\alpha\beta} \td \xi^\alpha \td \xi^\beta,
\end{equation} where $G_{\alpha \beta}$, $\alpha, \beta = \{0,1,2\}$ is a matrix satisfying $\det G =-\rho^2$ that encodes the inner products of the Killing vector fields $(\partial_t, \partial_{\phi^i})$ associated to the isometries and $\xi^i$ are coordinates adapted to their flows. The boundary set $\rho =0$ corresponds to the set of points upon which $G$ fails to have maximal rank; moreover as explained in considerable detail in \cite{Hollands:2007aj,Hollands:2008fm, Khuri:2017xsc} the $z-$axis then splits into a series of rods $I_i: z_{i} < z < z_{i+1}$ upon which either a timelike vector field becomes null (corresponding to a Killing horizon in spacetime) or a spacelike Killing vector field degenerates (corresponding to a symmetry axis in spacetime). Thus, each rod $I_i$ is characterized by an associated rod vector $v_i$ which denotes which linear combination of Killing fields is a null eigenvector of $G$. In particular, if the torus generators are chosen to have $2\pi-$periodic flows, the rod vector associated to a spatial rod must be an integer linear combination of the $\partial_{\phi^i}$. The collection of this data is referred to as the \emph{rod structure} of the solution. Additional constraints arising from the removal of conical and orbifold singularities can be found in \cite{Khuri:2017xsc}. The rod structure encodes the topology of the horizon and that of the asymptotic boundary at spatial infinity.

Remarkably, asymptotically flat black holes are uniquely characterized by their rod structure, mass,  and angular momenta \cite{Hollands:2007aj}. The question of existence of a solution given a particular rod structure can be addressed rigorously using the harmonic map structure underlying the Einstein equations in this setting \cite{Khuri:2017xsc,Khuri:2018udf}. For a given interior rod structure, asymptotically flat, asymptotically Kaluza-Klein, and asymptotically locally Euclidean (ALE) (namely, the spatial boundary at infinity is a lens space $L(p,q)$ with quartic volume growth) smooth black hole solutions can be shown to exist \cite{Khuri:2018udf}. The existence proof, however, uses abstract PDE techniques and the desired metric is not known fully explicitly; in particular, the analysis does not reveal whether conical singularities occur along the axes of symmetry. 

Explicit solutions can be produced using inverse scattering techniques \cite{Belinski:2001ph}, which are based upon the integrability of the stationary, biaxisymmetric vacuum Einstein equations. Remarkably, Chen and Teo have constructed explicit local families of black hole solutions that asymptotically approach gravitational solitons of the form \eqref{soliton} including the Kerr, Taub-NUT, and Eguchi-Hanson solitons \cite{Chen:2010ih}. This poses several open questions, including determining global properties of these spacetimes (e.g., investigating the possibility of causal pathologies, and whether smooth extensions exist through the event horizon) and the definition of conserved quantities associated to these non-asymptotically Minkowskian solutions. The purpose of the present work is to analyze in detail their solutions of the latter type, which can also be thought of as `ALE black holes' since, like the Eguchi-Hansonn soliton, their spatial slices have an asymptotic end diffeomorphic to $\mathbb{R} \times L(2,1)$.  We show that the metric can be extended smoothly across an event horizon with horizon topology $S^3$. We compute geometric invariants associated to this solution such as the mass and angular momenta, and show how these invariants can be computed using Komar integrals associated to spacetime symmetries or equivalently using an initial data perspective. We are particularly interested in how black hole mechanics is affected by the ALE asymptotics; we find that both a standard Smarr relation and variational first law hold. Finally, we compute the (finite) gravitational action of the associated complexified instanton using a mildly modified version of the renormalization prescription proposed by \cite{Kraus:1999di}. We find that the dominant contribution to the gravitational path integral with the ALE boundary conditions is a thermal Eguchi-Hanson soliton. 

This note is organized as follows. In Section 2 we present Chen-Teo's family of ALE black hole solutions and its rod structure (the latter is given in \cite{Chen:2010ih}, but with respect to a different basis).  We explicitly construct a Kruskal-Szekeres-type analytic extension through the event horizon and show that the domain of outer communications is stably causal. In Section 3 we compute geometric invariants and investigate thermodynamic properties of these black hole solutions. We conclude with a brief discussion in Section 4.
\section{The Chen-Teo family of solutions}
Chen and Teo present the following four-parameter family of five-dimensional, local Ricci flat metrics \cite{Chen:2010ih}:
 \begin{equation}
\begin{aligned}
\mathbf{g} &= -\frac{H(y,x)}{H(x,y)} (\td t - \omega_\psi \td \psi - \omega_\phi \td \phi)^2 - \frac{F(x,y)}{H(y,x)} \td \psi^2 + 2 \frac{J(x,y)}{H(y,x)} \td \psi \td\phi \\
& + \frac{F(y,x)}{H(y,x)} \td \phi^2 + \frac{\kappa^2 H(x,y)}{2(1-a^2)(1-b)^3 (x-y)^2} \left( \frac{\td x^2}{G(x)} - \frac{\td y^2}{G(y)} \right)
\end{aligned}
\end{equation} where $(t,y,x,\psi,\phi)$ are local coordinates. We will show below, after a global analysis, that $t$ can be interpreted as a timelike coordinate outside an event horizon, $(\psi,\phi)$ are coordinates on a torus with appropriate identifications required by regularity, and $(x,y)$ are coordinates on a closed rectangle. In summary, these local metrics are parameterized by four real constants $(\kappa, a , b , c)$ satisfying $\kappa > 0, -1 \leq a \leq 1, 0 \leq c \leq b <1$.  The ranges of the coordinates are $-\infty < t < \infty, -1 \le x \le 1, -1/c \le y \le -1$.  It turns out that a smooth solution arises provided $(a,b)$ are suitably chosen in terms of $c$ and $(\psi,\phi)$ are each independently periodic with period $2\pi$ (see below).  The metrics are of cohomogeneity-two, namely, they are invariant under the action of $\mathbb{R} \times \mathbb{T}^2$ generated by the Killing vector fields $\partial_t$ and $\partial_\psi, \partial_\phi$ respectively. The points $x = \pm 1$ and $y = -1$ correspond to axes (where some linear combination of $(\partial_\psi, \partial_\phi)$ degenerates) and $y = -1/c$ corresponds to the event horizon where a certain Killing field will become null. Spatial infinity corresponds to the region $x,y \to -1$ and cannot be described in the above coordinates; we will discuss the asymptotic region in detail below. This leaves the `scale' parameter $\kappa > 0$ and a parameter $c \in [0,1)$ that parametrizes the mass, angular momentum, and horizon area of the black hole. 
The functions appearing in the metric are complicated. For convenience, we present them here. 
\begin{align}
    G(x) &= (1 + c x ) (1-x^2) \\
    H(x,y) &= 4(1-b)(1-c)(1 + b x) \left[(1-b)(1-c) - a^2((1+bx)(1+cy) + (b-c)(1+y))\right] \nonumber \\
    & + a^2(b-c)(1+x)(1+y) \Bigl(1+b)(1+y)((1-a^2)(1-b)c(1+x) + 2 a^2 b(1-c)) \\
    & - 2b(1-b)(1-c)(1-x) \biggr) \nonumber \\
    F(x,y) & = \frac{2\kappa^2}{(1-a^2)(x-y)^2} \Bigl[4(1-c)^2(1+bx)[1 - b - a^2(1+bx)]^2 G(y) \\
    & - a^2G(x)(1+y)^2 \Bigl[1- b - a^2(1+b)^2]^2(1-c)^2(1+by) - (1-a^2)(1-b^2) \times \\
    & (1+cy) [(1-a^2)(b-c)(1+y) + [1 - 3 b - a^2(1+b)](1-c)]\biggr] \biggr] \\
    J(x,y) & = \frac{4 \kappa^2 a(1-c)(1+x)(1+y)}{(1-a^2)(x-y)} \left[1 - b - a^2(1+b)\right]\bigl((1-b)c + a^2(b-c)\biggr) \times \\
    & \left[(1+bx)(1+cy) + (1-cx)(1+by) + (b-c)(1-xy)\right].
\end{align} and 
\begin{equation}
\begin{aligned}
    \omega_\psi & = \frac{2\kappa}{H(y,x)} \left[\frac{2b(1+b)(b-c}{(1-a^2)(1+b)}\right]^{1/2}(1-c)(1+y) \Bigl(2[1-b - a^2(1+bx)]^2(1-c)  \\
    & -a^2(1-a^2)b (1-b)(1-x)(1+c x)(1+y) \biggr) \\
    \omega_\phi & = \frac{2\kappa}{H(y,x)}\left[\frac{2b(1+b)(b-c)}{(1-a^2)(1-b)}\right]^{1/2} a (1-c)(1+x)^2(1+y) \Bigl[a^4(1+b)(b-c) \\
    & + a^2(1-b)(-b + cb + 2c) - (1-b)^2 c \biggr].
\end{aligned}
\end{equation} Observe that the following identity holds: 
\begin{equation}
    \frac{F(x,y) F(y,x) + J(x,y)^2}{H(x,y)H(y,x)} = \frac{4 \kappa^4 G(x) G(y)}{(x-y)^4}.
\end{equation} This implies in particular that the left hand side must vanish at a $x = \pm 1, y = -1, y=-1/c$. 
\subsection{Rod structure}
In this coordinate system, 
\begin{equation}\label{eq:2.11}
-\det \mathbf{g} = \frac{\kappa^4 H(x,y)(F(x,y) F(y,x) + J(x,y)^2)}{4 (1-a^2)^2(1-b)^6 (x-y)^4 G(x) G(y) H(y,x)} = \frac{\kappa^8 H(x,y)^2}{ (1-a^2)^2(1-b)^6 (x-y)^8 }
\end{equation} The right hand side is positive\footnote{ Strict positivity of $H(x, y)$ is proved in Section 2.5} in the black hole exterior $y > -1/c$. Since the solution is stationary and bi-axisymmetric, the solution can be expressed in terms of the canonical Weyl-Pappapetrou coordinate system $(\rho,z)$ on the upper half plane $\rho >0, z \in \mathbb{R}$ with the boundary $\rho =0$ corresponding to fixed point sets of the spatial Killing vector fields and event horizon. In particular $\rho:=\sqrt{-\det G}$, where $G$ denotes the restriction of the spacetime metric $\mathbf{g}$ to the Killing vector fields $(\partial_t,\partial_\psi,\partial_\phi)$, is given by the positive root of 
\begin{equation}\label{rho}
\rho^2 = \frac{4\kappa^4(1-x^2)(y^2-1)(1 + c x)(1 + cy)}{(x-y)^4}.
\end{equation} We then have the following \emph{rod structure} where the axes of rotation are $x = \pm 1, y = -1$ and there is a null hypersurface where a timelike Killing field becomes null at $y = -1/c$. The spatial rod vectors $\ell_i$ given below are normalized to generate $2\pi-$periodic flows:
\begin{enumerate}
\item a semi-infinite rod at $x = -1$, $-1/c \le y  < -1$ with $\ell_1 = \partial_\phi$;
\item a timelike rod $y = -1/c$, $-1 \le x \le 1$ with associated Killing field 
\begin{equation}\label{xi}
\xi = \partial_t + \Omega_\psi \partial_\psi + \Omega_\phi \partial_\phi
\end{equation} where the angular velocities $\Omega_\psi, \Omega_\phi$ are given  by
\begin{equation}
\begin{aligned}
\Omega_\psi &= \frac{1}{\kappa(1-c)} \left[ \frac{(1-b)(b-c)}{2b(1-a^2)(1+b)}\right]^{1/2}, \\ \Omega_\phi &= \frac{a}{2\kappa} \cdot \frac{1 - b -a^2(1+b)}{(1-b)c + a^2(b-c)} \left[\frac{(1-b)(b-c)}{2b(1-a^2)(1+b)}\right]^{1/2}.
\end{aligned}
\end{equation}
\item a finite rod $x = 1$, $-1/c \le y \le-1$, with $\ell_2 = n \partial_\psi / \tilde{n} + \partial_\phi / \tilde{n}$ where
\begin{equation}
n = \frac{2a((1-b)c + a^2(b-c))}{ (1 - b - a^2(1+b))(1-c)}, \qquad \tilde{n} = \frac{(1-a^2)(1-b)(1+c)}{(1- b - a^2(1+b))(1-c)} \left[\frac{1-b}{1+b}\right]^{1/2} 
\end{equation} corresponding to a two-dimensional surface with the topology of a disc (hemisphere); 
\item a semi-infinite rod $y = -1$, $x \in (-1,1]$ and $\ell_3 = \partial_\psi$.
\end{enumerate} Observe that $\tilde{n} \ell_2 = n \ell_3 + \ell_1$. The choice corresponding to an ALE space with $L(2,1)$ topology at infinity is to take $\tilde{n} =2, n =1$. The absence of orbifold singularities at the corner points at the intersection of spatial rods imposes the conditions 
\begin{equation}\label{eq:2.16}
a = \frac{3(1-c)}{3 + 5c}, \qquad b = \frac{4c(3-c)}{5c^2 - 6c + 9}.
\end{equation} giving rise to a 2-parameter family of solutions $(\kappa,c$ with $0 < c < 1$ and $\kappa > 0$. We then have 
\begin{equation} \ell_3 = 2\ell_2 - \ell_1.
\end{equation} We now choose generators $K_1 := \ell_2$ and $K_2 := \ell_1$.  Define new angles $(\hat\phi^1, \hat \phi^2)$ adapted to the integral curves of $(K_1, K_2)$. As they each generate independent $2\pi-$periodic flows, the $(\hat\phi^1, \hat\phi^2)$ plane must have the following identifications: 
\begin{equation}
T_1: (\hat\phi^1, \hat \phi^2) \sim (\hat\phi^1 + 2\pi, \hat \phi^2), \qquad T_2: (\hat\phi^1, \hat \phi^2) \sim (\hat\phi^1, \hat \phi^2 + 2\pi)
\end{equation} These identifications can be expressed in terms of the $(\psi,\phi)$ coordinates: 
\begin{equation}\label{anglechange}
\frac{\partial}{\partial \hat \phi^1} = \frac{1}{2} \frac{\partial}{\partial \psi}+ \frac{1}{2}  \frac{\partial}{\partial \phi}, \qquad \frac{\partial}{\partial \hat \phi^2} =  \frac{\partial}{\partial \phi}
\end{equation} or equivalently
\begin{equation}
\psi = \frac{\hat\phi^1}{2} , \qquad \phi = \frac{\hat \phi^1}{2} + \hat \phi^2
\end{equation} and note $\td \psi \wedge \td \phi = \tfrac{1}{2} \td \hat \phi^1 \wedge \td \hat \phi^2$. The identifications $T_1, T_2$ are equivalent to \begin{equation}\label{ident1} (\psi, \phi) \sim (\psi + \pi, \phi + \pi), \qquad (\psi, \phi) \sim (\psi, \phi + 2\pi)
\end{equation} respectively. Applying $T_1$ twice and then $-T_2$ gives that $\psi, \phi$ are in fact independently $2\pi-$periodic, that is $\psi \sim \psi + 2\pi, \phi \sim \phi + 2\pi$; however $T_1, T_2$ are the minimal set. 

In summary, the rod structure is given in Figure \ref{rod} below.  The spatial rod vectors in the $(K_1, K_2)$ basis are $(0,1)$, $(1,0)$, and $(2,-1)$ (observe that they are integer-valued, as required). All the smoothness conditions are met, and from the rod structure we read off that spatial cross sections of the horizon have $S^3$ topology (since different Killing vector fields vanish on either side of the horizon rod). We will show below the spacetime metric can be extended smoothly through the event horizon. 
{
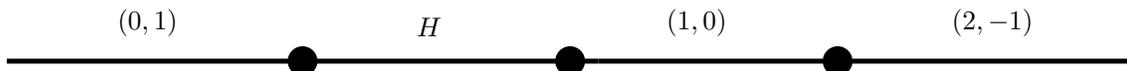
\begin{figure}[h!] 
\scalebox{1.56}{
\subfloat{
\begin{tikzpicture}[scale=1.2, every node/.style={scale=0.6}]
\draw[very thick](-4,0)--(-2,0)node[black,left=2cm,above=.2cm]{$(0,1)$};
\draw[very thick](-1.8,0)--(0.2,0)node[black,left=2.4cm,above=.2cm]{$H$};
\draw[very thick](0.2,0)--(1.8,0)node[black,left=1.8cm,above=.2cm]{$(1,0)$};
\draw[very thick](2,0)--(4,0)node[black,left=2cm,above=.2cm]{$(2,-1)$};
\draw[fill=black] (-1.9,0) circle [radius=.1] node[black,font=\large,below=.1cm]{};
\draw[fill=black] (0,0) circle [radius=.1] node[black,font=\large,below=.1cm]{};
\draw[fill=black] (1.9,0) circle [radius=.1] node[black,font=\large,below=.1cm]{};
\end{tikzpicture}} }
\caption{Rod structure for the Chen-Teo ALE black hole.}
\label{rod}
\end{figure}
\vspace{-1.5cm}
\subsection{Asymptotic geometry} In the $(x,y)$ coordinates, the orbit space can be described by a rectangle. This obscures the fact that the geometry contains an asymptotic region. This can be observed from \eqref{rho} by noting $\rho \to \infty$ as $(x,y) \to (-1,-1)$. We can better understand the geometry in this region by introducing polar-type coordinates $(r,\theta)$ via
\begin{equation}
r^2 =(1-c) \frac{\kappa^2 (2 - x - y)}{x-y}, \qquad \cos \theta = \frac{2 + x + y}{x-y}
\end{equation} These can be inverted to give
\begin{equation}
x = -\frac{r^2 - \kappa^2(2 + \cos\theta)(1-c)}{ r^2 +(1-c) \kappa^2 \cos\theta} , \qquad y = -\frac{ r^2 + (1-c)(2 - \cos\theta) \kappa^2}{r^2 + (1-c)\kappa^2 \cos\theta}
\end{equation} and then we want to find the expansions in $r$ of each metric function as $r \to \infty$. This yields
\begin{equation}\label{gttexpansion}
\mathbf{g}_{tt} = -1 + \frac{4b (1-c)\kappa^2}{(1-b) r^2} + O(r^{-4})
\end{equation}}
\begin{equation} \label{expansions}
\begin{aligned}
\mathbf{g}_{\psi\psi} & = r^2 \sin^2 (\frac{\theta}{2}) \\ &+ \kappa^2 \sin^2 (\frac{\theta}{2}) \frac{(1 + 3 b - 3 c - b c + a^2(-1 - 7b + 3c + 5bc) + 2(1-a^2)(1-b) c \cos \theta)}{(1-a^2)(1-b)} \\ & + O(r^{-2}) \\
\mathbf{g}_{\phi\phi} & =  \cos^2(\frac{\theta}{2})  r^2 \\
& + \kappa^2 \cos^2(\frac{\theta}{2})\left[ \frac{-1 + b + 3c - 3 b c + a^2(1 + 3b - 3c - b c) + 2(1 - a^2)(1-b) c \cos\theta}{(1-a^2)(1 - b)} \right] + O( r^{-2}) \\
\mathbf{g}_{\psi\phi} & =\frac{ 2 \sin^2\theta a \kappa^4 (1-c)}{\hat r^2(1-a^2)^2(1-b)^2} \cdot (a^2(b+1) + b-1)(a^2(b-c) + c(1 - b)) + O(\hat r^{-4})  \\
\mathbf{g}_{t\psi} & = -\frac{4 \kappa^3 (1-c) \sin^2(\theta/2)}{r^2} \left[ \frac{2 (1-a^2) b (1+b)(b-c)}{(1-b)^3}\right]^{1/2} + O(r^{-4}) \\
\mathbf{g}_{t\phi} & = O(r^{-6}) \\
\mathbf{g}_{r r} & = 1 + \frac{\kappa^2}{(1-a^2)(1-b) r^2} \left[2b(1-a^2)(1-c) + (1 + b - 5c + 3 bc + a^2(-1 + b(c-5) + 5c))\cos\theta \right] \\
& + O(\hat r^{-4}) \\
\mathbf{g}_{\theta\theta} & =  \left[ \frac{ r^2}{4} + \frac{(1-c)\kappa^2}{4(1-a^2)(1-b)} \left(2(1-a^2)b + (1 + b -a^2(1 + 5b))\cos\theta \right) \right] + O(r^{-2}) \\
\mathbf{g}_{r \theta} & = -\frac{c \kappa^2 \sin\theta}{r} + O(r^{-3})
\end{aligned}
\end{equation} Hence as $r \to \infty$ the geometry approaches the locally Minkowskian metric
\begin{equation}
    \td s^2 = -\td t^2 + \td r^2 + \frac{r^2}{4} \left(\td \theta^2 + \sin^2\left(\frac{\theta}{2}\right) \td \psi^2 + \cos^2\left(\frac{\theta}{2} \right) \td \phi^2 \right)
\end{equation} This metric is locally flat and the spatial level sets of $r$ have the local geometry of a round $S^3$. However, the identifications \eqref{ident1} imply that the globally these surfaces have lens space topology $L(2,1)$. One can introduce coordinates $\tilde \psi = \psi + \phi, \tilde\phi = \phi - \psi$ to transform the geometry to 
\begin{equation}
  \td s^2 =  -\td t^2 + \td r^2 + \frac{r^2}{4}\left(\td \theta^2 +(\td\tilde\psi + \cos\theta \td \tilde\phi)^2 + \sin^2\theta  \td \tilde\phi^2  \right)
\end{equation} where $(\tilde\psi, \tilde\phi) \sim (\tilde \psi + 2\pi, \tilde \phi)$, $(\tilde\psi, \tilde\phi) \sim (\tilde\psi + 2\pi, \tilde\phi + 2\pi)$ which exhibits the geometry of $L(2,1)$ as a $U(1)-$bundle over $S^2$ with Chern number 2 (to recover $S^3$ topology, $\tilde \psi$ would have to have period $4\pi$). Thus the Chen-Teo solution is asymptotic to a Minkowski spacetime quotiented by $\mathbb{Z}_2$. The latter spacetime is not a smooth manifold since it has an orbifold singularity at the origin. The appropriate `background' spacetime, identified by Chen and Teo, is the spacetime obtained by adding a trivial time direction to the Eguchi-Hanson gravitational instanton, which as is well known, is asymptotically locally Euclidean with asymptotic end $\mathbb{R} \times L(2,1)$. Indeed, when $c=0$, the timelike horizon rod shirnks to zero (this can be seen in the Weyl-Pappatrou coordinates). Setting 
\begin{equation}
    r = \left[\frac{\kappa^2 (2 - x -y)}{x-y}\right]^{1/2}, \qquad \cos \theta = \frac{2 + x + y}{x-y}
\end{equation} the full spacetime metric reduces to 
\begin{equation}\label{EH}
\td s^2 = -\td t^2 + \frac{\td r^2}{V(r)} + \frac{r^2}{4} \left[ V(r) (\td \tilde \psi + \cos\theta \td \tilde \phi)^2 + \td \theta^2 + \sin^2 \theta \td \tilde \phi^2 \right], \quad V(r) = 1 - \frac{\kappa^4}{r^4},
\end{equation} where $(\tilde\psi, \tilde\phi)$ are identified as above. The finite disc rod of the black hole solution becomes the familiar $S^2$ `bolt' at the origin $r = \kappa$ of the Eguchi-Hanson geometry. Hence it is natural to interpret the spacetime as describing an asymptotically Eguchi-Hanson spacetime containing an $S^3$ black hole. 
\subsection{Extension through the horizon}
As discussed above, the Killing vector field $\xi = \partial_t + \Omega_\psi \partial_\psi + \Omega_\phi \partial_\phi$ becomes null goes on the set $y = -1/c$. This is in fact a smooth Killing horizon generated by $\xi$. To see this, we can pass to a double null coordinate system and show that the metric can be extended analytically through the event horizon. Our strategy follows the approach used in constructing the maximal extension of the doubly-spinning black ring family \cite{Chrusciel:2009vr}.  We can introduce a coordinate system $(u,v,x,\Psi,\Phi)$ defined by 
\begin{equation}
    \td u = \td t + \frac{\alpha_1 \td y}{y - y_h} , \qquad \td v = \td t - \frac{\alpha_1 \td y}{y - y_h} 
\end{equation} and 
\begin{equation}
    \td \Psi = \td \psi - \alpha_2 \td t, \qquad \td \Phi = \td \phi - \alpha_3 \td t
\end{equation} where the $\alpha_i$ are constants. The new coordinate basis vectors are
\begin{gather}
    \partial_u = \frac{\partial_t}{2} + \frac{(y - y_h)}{2 \alpha_1} \partial_y + \frac{\alpha_2 \partial_\psi}{2} + \frac{\alpha_3 \partial_\phi}{2}, \qquad   \partial_v = \frac{\partial_t}{2} - \frac{(y - y_h)}{2 \alpha_1} \partial_y + \frac{\alpha_2 \partial_\psi}{2} + \frac{\alpha_3 \partial_\phi}{2} \\
    \partial_\Psi = \partial_\psi, \qquad \partial_\Phi = \partial_\phi.
    \end{gather}
We can then compute the components in the coordinate chart $(u,v,x,\psi,\phi)$ using $\mathbf{g}_{uu} = \mathbf{g}(\partial_u, \partial_u)$, $\mathbf{g}_{uv} = \mathbf{g}(\partial_u, \partial_v)$. In particular, $\mathbf{g}_{\Psi\Psi} = \mathbf{g}_{\psi \psi}$, $\mathbf{g}_{\Psi \Phi} = \mathbf{g}_{\psi\phi}$, $\mathbf{g}_{\Phi\Phi} = \mathbf{g}_{\phi\phi}$ and $\mathbf{g}_{xx}$ is unchanged. Let $\hat{H}:=H(y,x), H:=H(x,y)$. We compute
\begin{align}
    \mathbf{g}_{uv}  &= -\frac{(y-y_h)^2 \mathbf{g}_{yy}}{4\alpha_1^2} + \frac{1}{4}\left[\alpha_2^2 \mathbf{g}_{\psi\psi} + 2\alpha_2 \alpha_3 \mathbf{g}_{\psi\phi} + \alpha_3^2 \mathbf{g}_{\phi\phi}\right]  -\frac{\hat{H}}{4H} \left(1 - 2 \alpha_2\omega_\psi - 2 \alpha_3 \omega_\phi\right), \\
    \mathbf{g}_{uu} & = \frac{(y-y_h)^2 \mathbf{g}_{yy}}{4\alpha_1^2} +\frac{1}{4}\left[\alpha_2^2 \mathbf{g}_{\psi\psi} + 2\alpha_2 \alpha_3 \mathbf{g}_{\psi\phi} + \alpha_3^2 \mathbf{g}_{\phi\phi}\right]-\frac{\hat{H}}{4H} \left(1 - 2 \alpha_2\omega_\psi - 2 \alpha_3 \omega_\phi\right) = \mathbf{g}_{vv} \\
    \mathbf{g}_{u\Psi} & = \mathbf{g}_{v\Psi} = \frac{\hat{H}}{2H} \omega_\psi + \frac{1}{2} \left[\alpha_2 \mathbf{g}_{\psi\psi} + \alpha_3 \mathbf{g}_{\psi\phi}\right], \qquad \mathbf{g}_{u\Phi}  = \mathbf{g}_{v\Phi} = \frac{\hat{H}}{2H} \omega_\phi + \frac{1}{2} \left[\alpha_2 \mathbf{g}_{\psi\phi} + \alpha_3 \mathbf{g}_{\phi\phi}\right]
\end{align} Note that $(y-y_h)\mathbf{g}_{yy} = O(1)$ as $y \to y_h = -1/c$ and hence $(y-y_h)^2 \mathbf{g}_{yy} = O(y-y_h)$ and so the metric components are all analytic at $y = y_h$. Nonetheless, the metric in this coordintae system, which we will denote by $\mathbf{g'}$, is not invertible; the determinant of the metric in the new coordinate system is 
\begin{equation}
    \det \mathbf{g'} = \frac{(y- y_h)^2}{4\alpha_1^2} \det \mathbf{g} = -\frac{(y-y_h)^2\kappa^8 H(x,y)^2}{ 4\alpha_1^2 (1-a^2)^2(1-b)^6 (x-y)^8 }.
\end{equation} which vanishes on the event horizon.  To remedy this define new coordinates
\begin{equation}
    U = e^{\alpha_4 u}, \qquad V = e^{-\alpha_4 v}
\end{equation} where $\alpha_4$ is a constant. It follows that $\td U \wedge \td V = -\alpha_4^2 e^{\alpha_4(u-v)} \td u \wedge \td v$. Denoting the matrix representing the metric in the $(U,V,x,\Psi,\Phi)$ coordinates by $\mathbf{g''}$, 
\begin{equation}
    \det \mathbf{g''} = \frac{1}{\alpha_4^4} e^{-2\alpha_4(u-v)} \det \mathbf{g'}.
\end{equation} Note that 
\begin{equation}
    \td u - \td v = \td (u-v) = \frac{2\alpha_1 \td y}{y-y_h} \Rightarrow u - v = \int \frac{2\alpha_1 \td y}{y - y_h} = 2\alpha_1 \ln(y - y_h)
\end{equation} where an appropriate integration constant has been fixed, and hence 
\begin{equation}
    e^{-2\alpha_4(u-v)} = (y - y_h)^{-4 \alpha_4 \alpha_1}.
\end{equation} Choosing $\alpha_4 =  (2 \alpha_1)^{-1}$ then yields
\begin{equation}
    \det \mathbf{g''} = -\frac{4 \alpha_1^2 \kappa^8 H(x,y)^2}{(1-a)^2(1-b)^6 (x-y)^8}
\end{equation} which in particular does not vanish at the horizon. Note that
\begin{equation}
    U^2 = e^{2\alpha_4 t}(y- y_h), \qquad V^2 = e^{-2\alpha_4 t} (y-y_h)
\end{equation} The components of the metric in the $(U,V,x,\Psi, \Phi)$ chart are
\begin{equation}
    \mathbf{g}_{UU} = \frac{\mathbf{g}_{uu}}{\alpha_4^2 U^2}, \qquad \mathbf{g}_{VV} = \frac{\mathbf{g}_{vv}}{\alpha_4^2 V^2}, \qquad \mathbf{g}_{UV} = \frac{g_{uv}}{\alpha_4^2 UV}
\end{equation} and using the above expressions these simplify to
\begin{equation}
    \mathbf{g}_{UU} = \frac{e^{-2\alpha_4 t} g_{uu}}{\alpha_4^2 (y- y_h)}, \qquad \mathbf{g}_{VV} = \frac{e^{2\alpha_4 t} \mathbf{g}_{vv}}{\alpha_4^2 (y- y_h)}, \qquad \mathbf{g}_{UV} = \frac{\mathbf{g}_{uv}}{\alpha_4^2(y- y_h)}.
\end{equation} The cross terms of the form $\mathbf{g}_{U\phi^i}$ and $\mathbf{g}_{V \phi^i}$ with $\phi^i = (\Psi, \Phi)$ similarly involve the factor $(y- y_h)^{-1}$ multiplying $\mathbf{g}_{u \phi^i}, \mathbf{g}_{v\phi^i}$.  Regularity of the metric as expressed in the $(U,V,x,\Psi,\Phi)$ chart thus requires that the constants $(\alpha_1, \alpha_2, \alpha_3)$ be chosen so that $\mathbf{g}_{uu}= \mathbf{g}_{vv}, \mathbf{g}_{uv}, \mathbf{g}_{u\phi^i}=\mathbf{g}_{v\phi^i}$ vanish on the horizon. These are four conditions, and hence the system is overdetermined. 

We first consider the conditions $\mathbf{g}_{u\Psi} = \mathbf{g}_{u\Phi} =0$. These can be written in the form of a linear equation for the unknowns $(\alpha_2, \alpha_3)$
\begin{equation}
    \begin{pmatrix} A & B \\ B & C \end{pmatrix} \begin{pmatrix} \alpha_2 \\ \alpha_3 \end{pmatrix} = -\hat{H} \begin{pmatrix} \omega_\psi \\ \omega_\phi \end{pmatrix}
\end{equation} where $A, B, C$ are functions of $x$ given below,  
\begin{equation}
A = \frac{-FH}{\hat{H}} - \hat{H}\omega_\psi^2, \qquad
B = \frac{JH}{\hat{H}} - \hat{H}\omega_\psi\omega_\phi, \qquad
C = \frac{\hat{F}H}{\hat{H}} - \hat{H}\omega_\phi^2
\end{equation} and we have set $\hat F = F(y,x)$. 
The determinant of the matrix on the left hand side is
\begin{equation}
    AC - B^2 = -\frac{H^2}{\hat{H}^2} \left[F \hat{F} + J^2\right] + H \left[ 2 J \omega_\psi \omega_\phi - \hat{F} \omega_\psi^2 + F \omega_\phi^2 \right] = H F \left[ \omega_\phi + \frac{J \omega_\psi}{F} \right]^2
\end{equation} where we used the fact $\hat{F} F + J^2 =0$ on the horizon to simplify each term. We can solve the system: 
\begin{equation}
\begin{aligned}
    \alpha_2 &= \frac{-C\hat{H}\omega_\psi + B\hat{H}\omega_\phi}{AC - B^2} = -\frac{F(\hat{F} \omega_\psi - J \omega_\phi)}{\left[F \omega_\phi + J \omega_\psi\right]^2} = \frac{J}{F \omega_\phi + J  \omega_\psi} \\
    \alpha_3 & = \frac{B\hat{H}\omega_\psi - A\hat{H}\omega_\phi}{AC - B^2} = \frac{F(J\omega_\psi + F\omega_\phi)}{[J\omega_\psi + F\omega_\phi]^2} = \frac{F}{F\omega_\phi + J\omega_\psi}
\end{aligned}
\end{equation} 
Using \textit{Mathematica}, the expressions, evaluated at $y=y_h$, are constant as desired and take the form
\begin{equation}
\begin{aligned}
\alpha_2 &= \frac{(1-b) }{\sqrt{2} b (b+1) (1-c) \kappa }\sqrt{\frac{b (b+1) (b-c)}{\left(1-a^2\right) (1-b)}} ,\\
\alpha_3 &= \frac{a (b-1) \left(a^2 (b+1)+b-1\right) }{2 \sqrt{2} b (b+1) \kappa  \left(-c
   \left(a^2+b\right)+a^2 b+c\right)}\sqrt{\frac{b (b+1) (b-c)}{\left(1-a^2\right) (1-b)}}.
\end{aligned}
\end{equation}
Next, a similar \textit{Mathematica} calculation shows that $\mathbf{g}_{uu}$ will have a second order zero at $y = y_h$ and thus make $\mathbf{g}_{UU}$ regular if and only if the constant $\alpha_1$ satisfies
\begin{equation}
\begin{aligned}
\alpha_1 &= \sqrt{2} \sqrt{\frac{b (b+1) (1-c)^2 \kappa ^2 \left(-c \left(a^2+b\right)+a^2 b+c\right)^2}{\left(1 - a^2\right) (1-b)^4 c (c+1)^2}} .
\end{aligned}
\end{equation}
Using these constants, we see that the chart $(U,V,x,\Psi,\Phi)$ the metrix ia regular at $U=0$ or $V=0$ and can be continued from the region $U,V \geq 0$ to $U, V < 0$. 

\subsection{Stable causality} A black hole spacetime should satisfy certain global regularity conditions. In particular, the black hole exterior should be globally hyperbolic. The aim of this section is to prove analytically that the spacetime is stably causal, that is, there exists a global function whose gradient is everywhere timelike and future directed. This allows one to unambiguously distinguish between the future and past of each point in the exterior. The natural choice for a time function is the coordinate $t$ with $\mathbf{g}^{tt} = \mathbf{g}^{-1}(\nabla t, \nabla t)$. We see that from the form of the metric and identity (\ref{eq:2.11}) that
\begin{equation}
\begin{aligned}
\mathbf{g}^{tt} & = \frac{\mathbf{g}_{xx}\mathbf{g}_{yy}}{\det \mathbf{\mathbf{g}}}\det \begin{pmatrix}
\mathbf{g}_{\psi\psi} & \mathbf{g}_{\psi\phi}\\\mathbf{g}_{\psi\phi} &  \mathbf{g}_{\phi\phi}
\end{pmatrix} \\& =  -\frac{(x-y)^4}{4\kappa^4 G(x)G(y)}\det\begin{pmatrix}
\mathbf{g}_{\psi\psi} & \mathbf{g}_{\psi\phi}\\g_{\psi\phi} &  \mathbf{g}_{\phi\phi}
\end{pmatrix}
\\ & = \frac{\Theta_1(x, y, a, b, c)}{\Theta_2(x, y, a, b, c)}
\end{aligned}
\end{equation}
where $\Theta_1, \Theta_2$ are polynomial functions in the coordinates $x, y$ and depend on the parameters $a, b, c$. It is difficult to check the sign for general parameters $a, b$, due to a potential root in both the numerator and  denominator terms. However, for the Chen-Teo solution which gives rise to the Eguchi-Hanson metric no such root appears, and we can proceed. We use the approach used in \cite{Chrusciel:2009vr}.

Using (\ref{eq:2.16}), we can rewrite the polynomials only in terms of parameter $c$ 
$$\mathbf{g}^{tt} = \frac{P_1(x, y, c)}{P_2(x, y, c)}$$
The following change of variables can be used to show that each $P_1, P_2$ has a sign: let $\eta \in [0, \infty)$, $\gamma \in (0, \infty)$ and $\lambda \in [0, \infty)$ with redefinition
\begin{equation}\label{eq:2.53}
x = -1 + \frac{2}{1 + \eta} \qquad c = \frac{1}{1 + \gamma} \qquad y = -\frac{1 + \lambda + \gamma \lambda}{1 + \lambda}
\end{equation}
which leads to the right ranges except for $x = -1$ or $c = 0$  which we consider later. 
\\\\
Inserting the above substitutions into $\mathbf{g}^{tt}$ one obtains
\begin{equation}
\begin{aligned}
P_1(\eta, \gamma, \lambda) = -\frac{Q(\eta, \gamma, \lambda)}{(1 + \eta)^3(1 + \gamma)^3(8 + 3\gamma)^3(8 + 12\gamma + 9\gamma^2)^4(1 + \lambda)^4}
\\
 P_2(\eta, \gamma, \lambda) = \frac{R(\eta, \gamma, \lambda)}{(1 + \eta)^2(1 + \gamma)^3(8 + 3\gamma)^3(8 + 12\gamma + 9\gamma^2)^4(1 + \lambda)^4}
\end{aligned}
\end{equation}
Here $Q, R$ are polynomials in $(\eta, \gamma, \lambda)$.  A \textit{Mathematica} calculation shows that all coefficients for $Q$ (up to an overall scaling) are in the set $[944784, 7419174912] \cap \mathbb{Z}$ and moreover that $Q \ge 63700992\gamma^7$, therefore this proves strict positivity for $Q$. A similar calculation shows that all numerical coeffecients for $R$ are in the set $[944784, 636014592] \cap \mathbb{Z}$ and moreover that $R \ge 35831808\gamma^9$, therefore this proves strict positivity for $R$. 

We now know that strict negativity of $\mathbf{g}^{tt}$ is true in the region
\begin{equation}
\Xi = \{(x, y, c) : -1 < x \le 1, -\frac{1}{c} < y \le -1, 0 < c < 1\}
\end{equation}
Therefore we can now consider the boundary of $\Xi$ now. For the case where $x = -1$ and $c \ne 0$, proceed as before except set $x = -1$ in $P_1, P_2$. Substitution yields two polynomials
\begin{equation}
\begin{aligned}
P_1(\gamma, \lambda) = -\frac{Q_x(\gamma, \lambda)}{(1 + \gamma)(8 + 3\gamma)(8 + 12\gamma + 9\gamma^2)^3(1 + \lambda)^3}\\
P_2(\gamma, \lambda) = \frac{729\gamma^7(2 + (2 + \gamma)\lambda)}{(1 + \gamma)(8 + 12\gamma + 9\gamma^2)^3(1 + \gamma)^2}
\end{aligned}
\end{equation} where $Q_x$ is a polynomial in $(\gamma, \lambda)$ A \textit{Mathematica} calculation shows that up to an overall scaling, all the numerical coefficients for $Q_x$ are in the set $[2187, 67392] \cap \mathbb{Z}$ and moreover that $Q_x \ge 11664\gamma^7$, therefore this proves strict positivity for $Q_x$. We can also see that $P_2$ is strictly positive as well. We now know that strict negativity of $\mathbf{g}^{tt}$ is true in the part of the boundary
\begin{equation}
\{(x, y, c) : x = -1, -\frac{1}{c} < y \le -1, 0 < c < 1\} \subset \partial \Xi
\end{equation}
\\
The remaining case is for $c = 0$ for which we note that the metric reduces to $\mathbf{g}^{tt} = -1$ which is always trivially negative. Therefore we have proved stable causality in the domain of outer communications for the Eguchi-Hanson case of the Chen-Teo black hole solutions.

\subsection{Positivity of the function $H(x,y)$.}
Regularity of the metric clearly requires that we prove strict positivity for $H(x, y)$. Consider the regions for $c \ne 0$,
\begin{equation}
\begin{aligned}
\Xi_1 &= \{-1 < x \le 1, -1/c < y \le -1,\; c > 0\},\\
\Xi_2 &= \{x = -1, -1/c < y \le -1,\; c > 0\},\\
\Xi_3 &= \{y = -1/c, -1 < x \le 1,\; c > 0\},\\
\Xi_4 &= \{x = -1, y = -1/c,\; c > 0\}.
\end{aligned}
\end{equation}
Using the same techniques in the previous section we re-express these polynomials in terms of variables $\eta$, $\gamma$, $\lambda$ as in (\ref{eq:2.53}). Then the resulting expressions become 
\begin{equation}
\begin{aligned}
H_1(\eta, \gamma, \lambda) &= \frac{Q_{H_1}(\eta, \gamma, \lambda)}{(1 + \eta)^2(1 + \gamma)^2(8 + 3\gamma)^3(8 + 12\gamma + 9\gamma^2)^3(1 + \lambda)^2}\\
H_2(\gamma, \lambda) &= \frac{11664 \gamma ^8 \left(3 \gamma ^2 \lambda +2 \gamma  (7 \lambda +6)+16 (\lambda
   +1)\right)}{(\gamma +1)^2 (3 \gamma +8)^2 \left(9 \gamma ^2+12 \gamma +8\right)^3
   (\lambda +1)}\\
H_3(\eta, \gamma) &= \frac{Q_{H_3}(\eta, \gamma)}{(\eta +1)^2 (\gamma +1)^2 (3
   \gamma +8)^3 \left(9 \gamma ^2+12 \gamma +8\right)^3}\\
H_4(\gamma) &= \frac{11664 \gamma ^8 (\gamma +2)}{(\gamma +1)^2 (3 \gamma +8) \left(9 \gamma ^2+12 \gamma
   +8\right)^3}
\end{aligned}
\end{equation}
Here $Q_{H_1}, Q_{H_3}$ are polynomials in a subset of $(\eta, \gamma, \lambda)$. A \textit{Mathematica} calculation shows that all the coefficients for $Q_{H_1}, Q_{H_2}$ are in the set $[104976, 17915904] \cap \mathbb{Z}$. Moreover, we see that $Q_{H_1} \ge 1990656\gamma^6$ and $Q_{H_3} \ge 2654208\gamma^6$, therefore this shows that $H(x, y)$ is strictly positive for $\Xi_1 \cup \Xi_2 \cup \Xi_3 \cup \Xi_4 = \{-1 \le x \le 1, -1/c \le y \le 1\}$ where $c > 0$ (when $c = 0$, the $(x,y)$ coordinates are degenerate and the solution reduces to the Eguchi-Hanson soliton).
\subsection{Ergosurfaces and ergoregions}
\begin{figure}[h]
    \centering
    \includegraphics[width=0.32\linewidth]{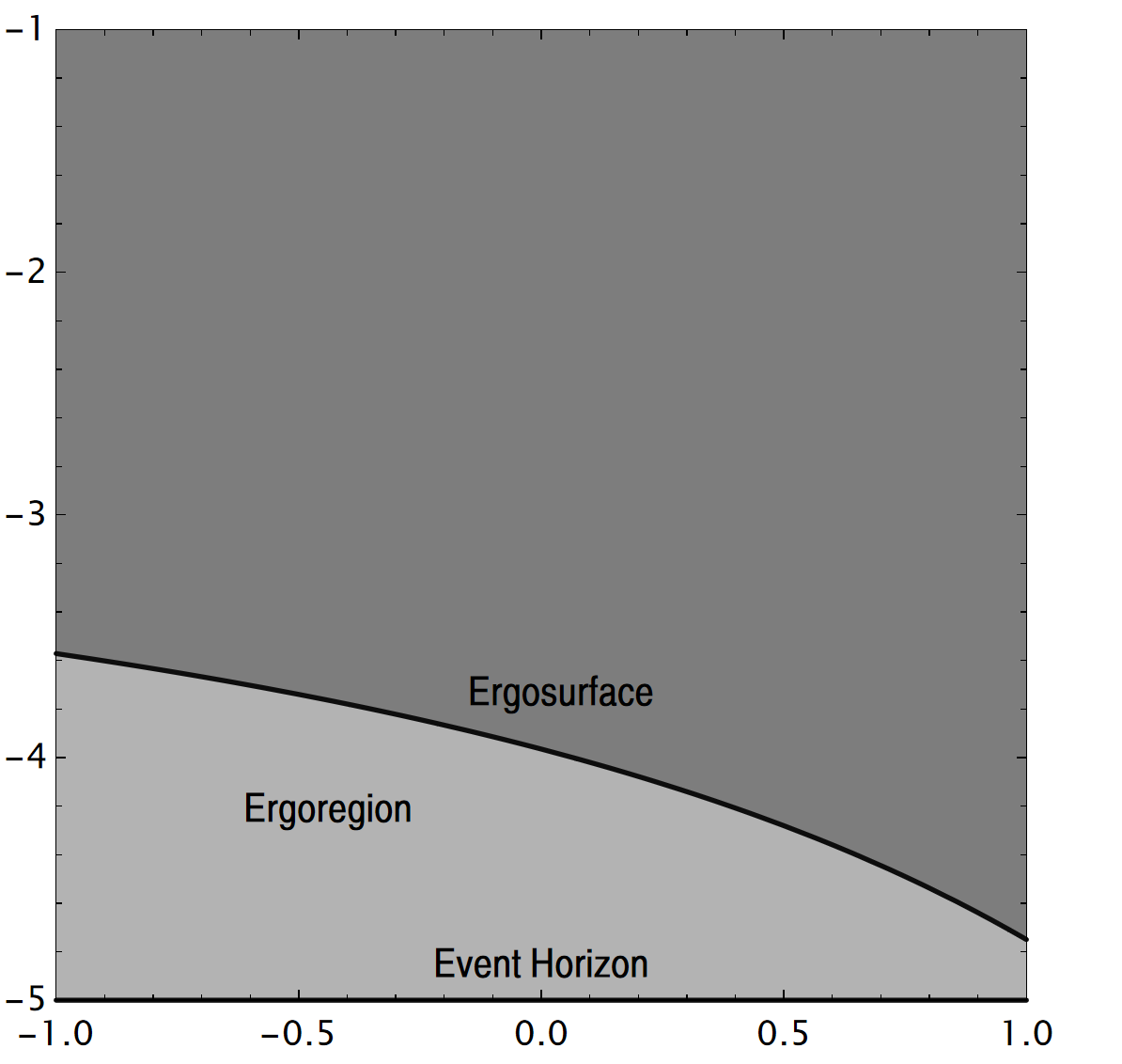}\hfill
    \includegraphics[width=0.33\linewidth]{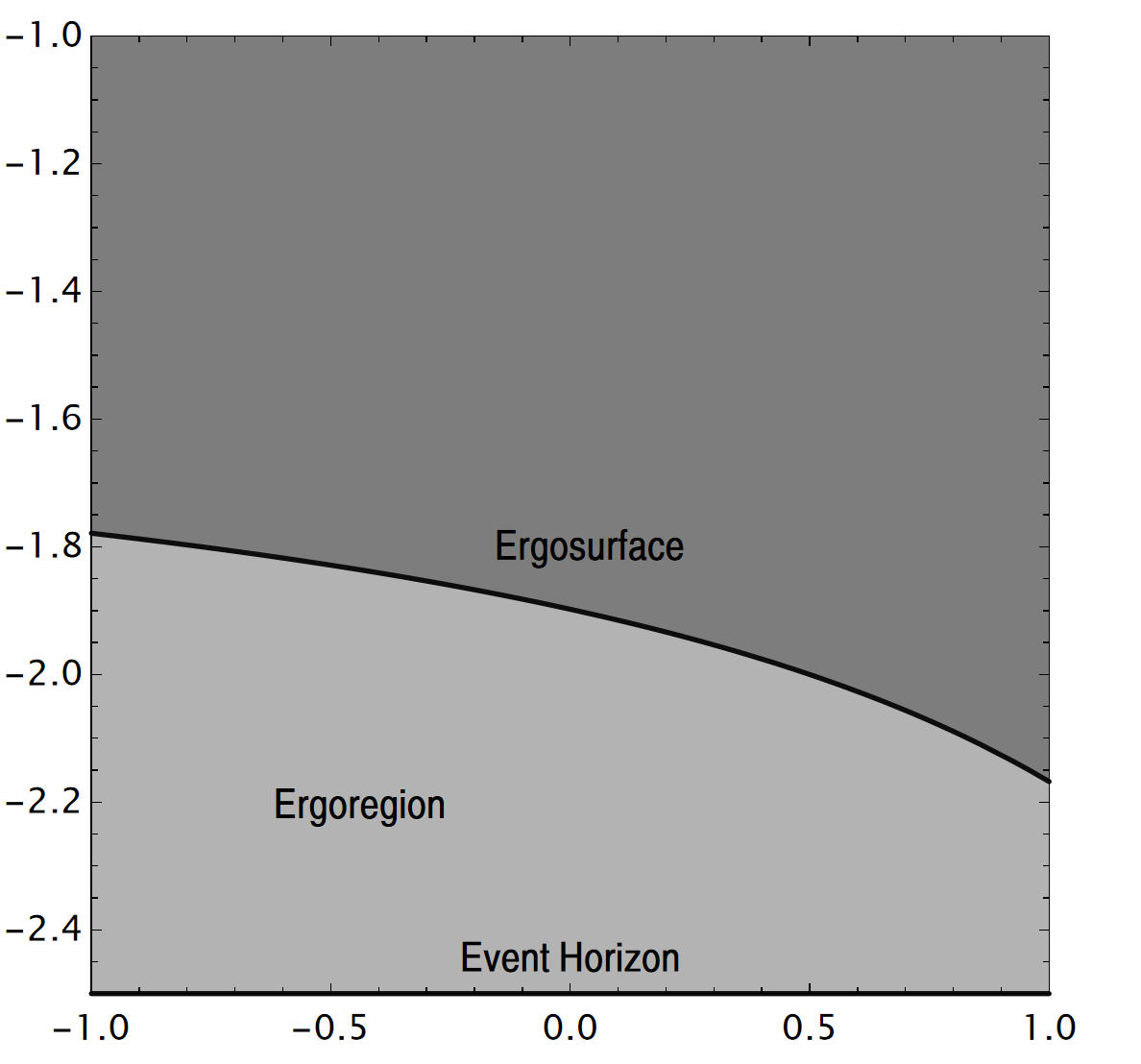}
    \includegraphics[width=0.34\linewidth]{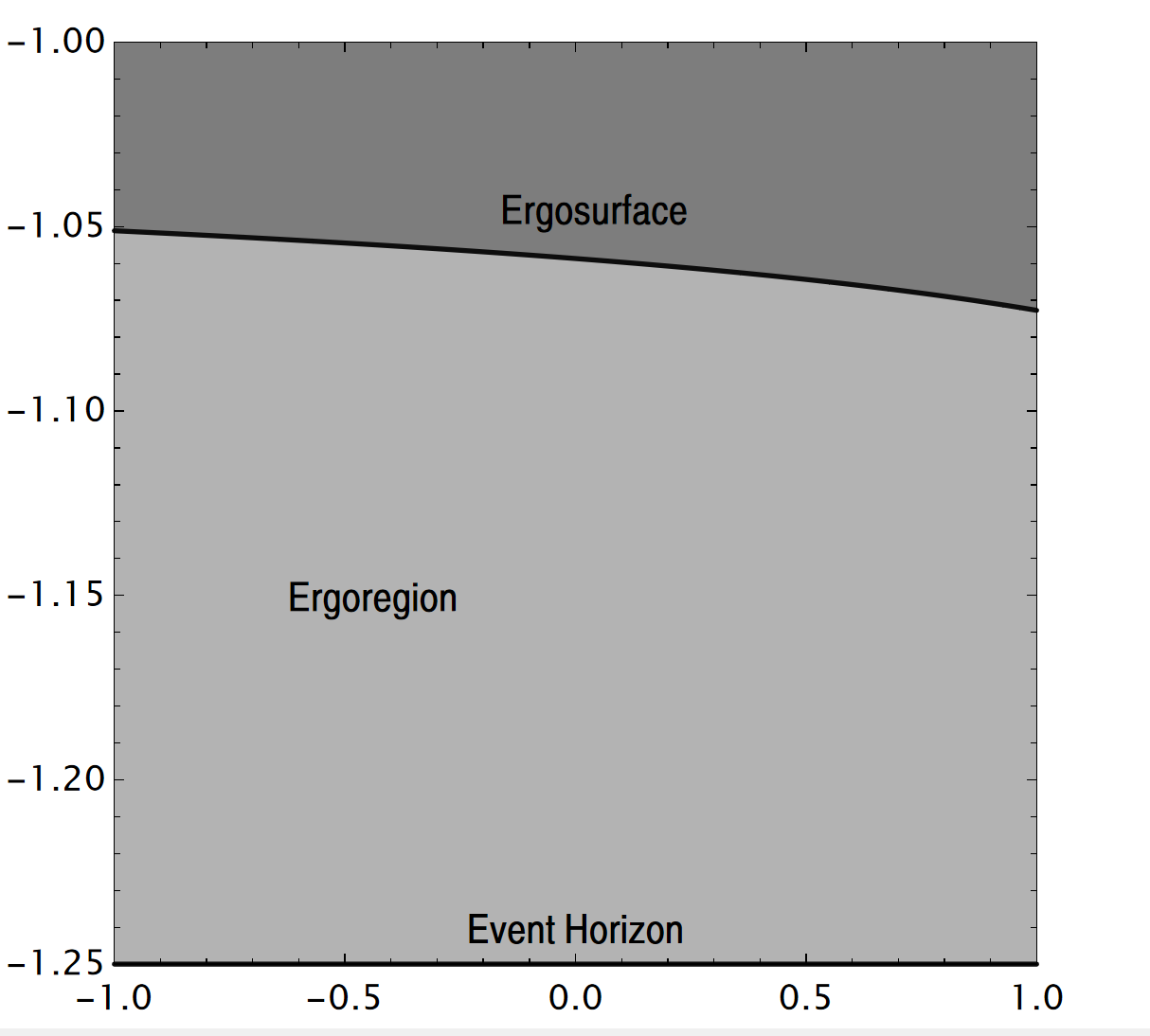}
    \caption{Ergoregion in the $(x,y)$ plane for $c = (0.2, 0.4, 0.8)$.}
    \label{fig:ergo}
\end{figure}

Ergoregions will occur in spacetime regions when the asymptotically stationary Killing field $\partial_t$  is spacelike, i.e. $\mathbf{g}_{tt} > 0$. The boundary of this region is the locus of points where $\mathbf{g}_{tt}=0$ which reduces to the implicit equation $H(y,x)=0$. This requirement is independent of the scale parameter $\kappa$ and hence depends only on $c$. It can be shown analytically that for all $c \in (0,1)$, the ergosurface does not intersect the horizon, that is, there are no solutions to $H(-1/c,x)=0$ for $x \in [-1,1]$ (see Figure \ref{fig:ergo}). This is in contrast to the familiar situation in Kerr. Moreover, plotting the ergosurface shows that it is connected. Finally, as $c \rightarrow 1$, we see that the mass and angular momentum of the black hole increase for given $\kappa$ and the size of the ergoregion grows in the $(x,y)$ domain. 

\section{Physical properties and thermodynamics} In this section we will investigate some physical and geometric invariants associated to this smooth black hole spacetime, with particular emphasis on variational identifies (i.e. `the first law') satisfied by nearly members in this family. 
\subsection{Geometry of the horizon}
The geometry of spatial cross sections of the horizon is given by 
\begin{equation}
\begin{aligned}
g_H&= -\frac{H(y_h,x)}{H(x,y_h)} (\omega_\psi \td \psi + \omega_\phi \td \phi)^2 - \frac{F(x,y_h)}{H(y_h,x)} \td \psi^2 + 2 \frac{J(x,y_h)}{H(y_h,x)} \td \psi \td\phi \\
& + \frac{F(y_h,x)}{H(y_h,x)} \td \phi^2 + \frac{\kappa^2 H(x,y_h)}{2(1-a^2)(1-b)^3 (x-y_h)^2} \cdot \frac{\td x^2}{G(x)} 
\end{aligned}
\end{equation} where $y_h = -1/c < 0$.  According to the rod structure (see aove), this is extends to an inhomogeneous metric on $S^3$. The determinant of the metric in the $(x,\psi,\phi)$ coordinates, is 
\begin{equation}
\det g_H =\kappa^6 \cdot \frac{8 b (1+b)(1 - c)^4 c (a^2(b-c) + c(1-b))^2}{(1-a^2)(1-b)^4 (1 + c x)^4}
\end{equation} We can then compute the area of the horizon to find
\begin{equation}\label{A_H}
A_H  = 2\pi^2 \int_{-1}^1 \sqrt{\det g_H} \td x = \frac{8 \pi^2 \kappa^3 (1-c) (a^2(b-c) + (1-b) c )}{(1-b)^2 (1+c)} \cdot \left(\frac{ 2bc(1+b) }{1-a^2} \right)^{1/2}
\end{equation} where we used $\td \psi \wedge \td \phi = \tfrac{1}{2} \td \hat \phi^1 \wedge \hat \td \phi^2$ where the angles $\hat \phi^i$ each parameterize $2\pi$-periodic circles.  The surface gravity $\kappa_g$, defined by $\nabla_\xi \xi = -2\kappa_g \xi$ evaluated on the horizon (recall that the null generator is given by \eqref{xi}) is found to be
\begin{equation}
\kappa_g = \frac{1}{4\kappa (1-c)} \frac{(1-b)^2(1+c)}{(a^2(b-c) + (1-b)c} \cdot \left(\frac{2 c (1-a^2)}{b(1+b)} \right)^{1/2}. 
\end{equation} Observe that there is no extreme limit within the allowed range of parameters $c \in (0,1)$; the event horizon is always non-degenerate. The situation is analogous to that of a singly-spinning Myers-Perry black hole. 
\subsection{Mass and angular momenta}
To compute the ADM mass we need to pick an ALE spatial slice $\Sigma$. The easiest choice is just $t =$constant. In the language of dynamics, we can think of this time slice as an initial data set $(\Sigma, g, K)$. The Riemannian metric induced on the hypersurface $t =$constant is given by
 \begin{equation}
\begin{aligned}\label{slicemet}
g = g_{ij} \td x^i \td x^j &= -\frac{H(y,x)}{H(x,y)} ( \omega_\psi \td \psi + \omega_\phi \td \phi)^2 - \frac{F(x,y)}{H(y,x)} \td \psi^2 + 2 \frac{J(x,y)}{H(y,x)} \td \psi \td\phi \\
& + \frac{F(y,x)}{H(y,x)} \td \phi^2 + \frac{\kappa^2 H(x,y)}{2(1-a^2)(1-b)^3 (x-y)^2} \left( \frac{\td x^2}{G(x)} - \frac{\td y^2}{G(y)} \right)
\end{aligned}
\end{equation} and $x^i = (y,x,\psi, \phi)$. For convenience we will denote the spatial metric by $g$ and reserve $\mathbf{g}$ for the spacetime metric. To compute the 2nd fundamental form $K_{ij}$ we use the ADM decomposition of the spacetime metric
\begin{equation}\label{ADMLor}
\mathbf{g} = -\alpha^2 \td t^2 + g_{ij}(\td x^i  + N^i \td t)(\td x^j + N^j \td t) 
\end{equation} where $\alpha > 0$ and $N^i$ are the lapse and shift respectively.  Note that 
\begin{equation}
\mathbf{g}_{tt} = -\alpha^2 + g_{ij} N^i N^j, \qquad \mathbf{g}_{ti} = g_{ij}N^j . 
\end{equation} We treat $N^i$ as a vector field on $(\Sigma, g)$ and raise and lower indices appropriately with $g^{ij}$ and $g_{ij}$ respectively. The 2nd fundamental form for a stationary metric is given by
\begin{equation}
K_{ij} = (2\alpha)^{-1} (D_i N_j + D_j N_i) 
\end{equation} which is defined only up to an overall sign and $D$ is the metric connection on $(\Sigma, g)$. One finds that the only non-vanishing components of the 2nd fundamental form are
\begin{equation}\label{extcurv}
K_{aI} = \frac{1}{2\alpha} \left(g_{IK} \partial_a N^K \right)
\end{equation} where the indices $A=(x,y)$ and $I= (1,2)$ denote angular coordinates $(\psi, \phi)$ associated to the Killing vector fields $\partial_\psi, \partial_\phi$.  This shows that the initial data set is `$t-\phi^i$' symmetric (i.e. the spacetime metric is invariant under the discrete isometry $t \to - t, \psi \to -\psi, \phi \to - \phi$. This implies that the initial data set is maximal, that is $\text{Tr}_g K=0$. This can of course be verified in this particular case by computing the trace of \eqref{extcurv}. Note that $(\Sigma, g_{ij}, K_{ij})$ constitute an ALE initial data set for the vacuum Einstein equations, and hence satisfy
\begin{equation}
    R(g) - |K|^2_g = 0, \qquad \text{div}_g K =0.
\end{equation} where $R(g)$ is the scalar curvature of $g$. 

To compute the ADM quantities, it is useful to compute the asymptotic behaviour of the metric in an asymptotically Cartesian coordinate chart. We already have the expansions above in the natural coordinates $(r, \theta, \psi, \phi)$ in \eqref{expansions}. To pass to Cartesian coordinates one needs
\begin{equation}\begin{aligned}
x_1 &= r \sin (\theta/2) \cos \psi, \quad x_2 = r \sin(\theta/2) \sin\psi, \\ x_3 &= r \cos(\theta/2) \cos \phi, \quad x_4 = r \cos (\theta/2) \sin \phi . 
\end{aligned}
\end{equation} which can be inverted: 
\begin{align}
r &= \sqrt{x_1^2 + x_2^2 + x_3^2 + x_4^2}, \qquad
\theta = 2\arctan \left[ \frac{x_1^2 + x_2^2}{x_3^2 + x_4^2} \right]^{1/2}, \\
\psi &= \arctan \left[\frac{x_2}{x_1}\right] \qquad \phi  = \arctan \left[\frac{x_4}{x_3}\right]
\end{align} which implies
\begin{equation}
\td \psi = \frac{x_1 \td x_2 - x_2 \td x_1}{x_1^2 + x_2^2}, \qquad \td \phi = \frac{x_3 \td x_4 - x_4 \td x_3}{x_3^2 + x_4^2}
\end{equation} and 
\begin{equation}\begin{aligned}
\td r & = \frac{1}{ r} ( x_1 \td x_1 + x_2 \td x_2 + x_3 \td x_3 + x_4 \td x_4) \\
\td \theta & = \frac{2}{ r^2} \left[\left(\frac{x_3^2 + x_4^2}{x_1^2 + x_2^2}\right)^{1/2} x_1 \td x_1 + \left(\frac{x_3^2 + x_4^2}{x_1^2 + x_2^2}\right)^{1/2}  x_2  \td x_2 - \left(\frac{x_1^2 + x_2^2}{x_3^2 + x_4^2}\right)^{1/2}  x_3 \td x_3 \right. \\ 
&\left. - \left(\frac{x_1^2 + x_2^2}{x_3^2 + x_4^2}\right)^{1/2} x_4 \td x_4 \right]
\end{aligned}
\end{equation} Now if we write $g_{\psi\psi} = r^2 \sin^2(\theta/2) + c_\psi, g_{\phi\phi} = r^2 \cos^2(\theta/2) + c_\phi, g_{\psi\phi} = c_{\psi\phi} /r^2, g_{r r} = 1 + c_r / r^2, g_{\theta\theta} = r^2/4 + c_\theta, g_{r \theta} = \frac{c_{r \theta}}{r}$,  then as $r \to \infty$,
\begin{equation}
g \to \delta_4 + \frac{c_r \td \hat r^2}{\hat{r}^2} + c_\theta \td \theta^2 + \frac{2c_{\theta \hat r}}{\hat r} \td \hat r \td \theta + c_\psi \td \psi^2 + c_\phi \td \phi^2 + 2 \frac{c_{\psi\phi} }{\hat r^2}\td \psi \td \phi
\end{equation} where $\delta_4$ is the flat Euclidean metric and the constants $c_r, c_\theta, c_{r\theta}, c_\psi, c_\phi, c_{\psi\phi}$ are given by
\begin{equation}
\begin{aligned}
c_r &= \frac{\kappa ^2 \left(\cos (\theta ) \left(a^2 ((b+5) c-5 b-1)+3 b c+b-5 c+1\right)+2 \left(a^2-1\right) b (c-1)\right)}{\left(a^2-1\right) (b-1)}\\
    c_\theta &= \frac{(c-1) \kappa ^2 \left(\left(a^2 (5 b+1)-b-1\right) \cos (\theta )+2 \left(a^2-1\right) b\right)}{4 \left(a^2-1\right) (b-1)}\\
    c_{r\theta} &= -{c \kappa ^2 \sin (\theta )}\\
    c_{\psi} &= \frac{\kappa ^2 \sin ^2\left(\frac{\theta }{2}\right) \left(2 \left(a^2-1\right) (b-1) c \cos (\theta )+a^2 (5 b c-7 b+3 c-1)-(b+3) c+3 b+1\right)}{\left(a^2-1\right) (b-1)}\\
    c_{\phi} &= \frac{\kappa ^2 \cos ^2\left(\frac{\theta }{2}\right) \left(2 \left(a^2-1\right) (b-1) c \cos (\theta )+a^2 (-(b+3) c+3 b+1)-3 b c+b+3 c-1\right)}{\left(a^2-1\right) (b-1)}\\
    c_{\psi\phi} &= \frac{2 a (c-1) \kappa ^4 \left(a^2 (b+1)+b-1\right) (\cos (\theta )-1) (\cos (\theta )+1) \left(-c \left(a^2+b\right)+a^2 b+c\right)}{\left(a^2-1\right)^2 (b-1)^2}
\end{aligned}
\end{equation}
Recall that the ADM mass is
\begin{equation}
m_{ADM} = \frac{1}{16\pi} \int_{S_\infty} \left(n^i [\partial^j h_{ij} - \partial_i \tr h ]\right) \td \text{Vol} 
\end{equation} where here $S_\infty$ is the asymptotic boundary metric on a level set $r$ as $r \to \infty$ of a constant time slice and $n = \partial_r$ is the unit normal to the level set.  In this case that space will be the lens space $L(2,1)$ and note $\td \text{Vol} = \tfrac{1}{4}r^3 \sin(\theta/2) \cos(\theta/2) \td \theta \wedge \td \hat\phi^1 \wedge \td \hat \phi^2$ where each $\hat \phi^i$ has period $2\pi$ and $0 < \theta < \pi$.  The trivial integration over the angles $(\hat \phi^1, \hat \phi^2)$ gives a factor of $4\pi^2$. A computation gives
   \begin{equation}
   m_{ADM} = \frac{ 3 b (1 - c) \pi \kappa^2}{4(1-b)}.
   \end{equation}
Note that this is indeed proportional to the $O(1/\hat{r}^2)$ term in the expansion of $\mathbf{g}_{tt}$ given above.  Defining
\begin{equation}
m_{K} := -\frac{3}{32\pi } \int_{S} \star \td K 
\end{equation} with asymptotic boundary at infinity here corresponding to $L(2,1)$, we find 
\begin{equation}
    m_K =  \lim_{\hat r\to \infty} \frac{3\pi}{32} (-g_{tt}'(\hat{r})) r^3 = m_{ADM}
\end{equation} which fixes the normalization for the Kormar integral.  Explicitly,$K = \partial_t$ and
\begin{equation}
    \star \td K\vert_{S} = \left[ \partial_r \mathbf{g}_{tt} (-\mathbf{g}^{tt}\mathbf{g}^{rr}) + \partial_r \mathbf{g}_{t\psi}(-\mathbf{g}^{tr} \mathbf{g}^{\psi t}) + \partial_r \mathbf{g}_{t\phi} (-\mathbf{g}^{rr} \mathbf{g}^{t\phi})\right] \td \text{Vol}
\end{equation}

To compute the angular momenta, we must compute the asymptotic behaviour of the second fundamental form using \eqref{extcurv}. It is easily checked that $\alpha = 1 + O(1/r^2)$ as $r \to \infty$. One finds
\begin{equation}
K_{r\psi} = \frac{8 \sqrt{2}(1-c) \kappa^3}{1-b} \sin^2 (\theta/2) \left[ \frac{b(1+b)(b-c)(1-a^2)}{1-b}\right]^{1/2} r^{-3} + O(r^{-5}) , \quad K_{r \phi} = O(r^{-7}),
\end{equation} which yields
\begin{equation}
J_\psi =  \frac{\pi \sqrt{2}(1-c) \kappa^3}{2(1-b)}  \left[ \frac{b(1+b)(b-c)(1-a^2)}{1-b}\right]^{1/2} , \qquad J_\phi =0.
\end{equation} Observe that $J_\phi =0$ although $\Omega_\phi \neq 0$. We would like to compare this with the Komar type expression for the angular momenta, 
\begin{equation}
J^K_i = \frac{c_1}{16 \pi} \int_{S_\infty} \star \td \eta_{(i)}
\end{equation} where $c_1$ is a constant to be chosen so that $J^K_i = J_i$. 
It is useful to have the expansions for the inverse metric components:
\begin{align}
\mathbf{g}^{tt} & = -1 - \frac{4\kappa^2 b (1-c)}{(1-b) r^2} + O(r^{-4}) \\
\mathbf{g}^{t\psi} & =- \frac{4\kappa^3 \sqrt{2} (1-c)}{(1-b)r^4} \left[\frac{b (1+b)(b-c)(1-a^2)}{1-b}\right]^{1/2} + O(r^{-6}), \quad \mathbf{g}^{t\phi} = O(r^{-8}) \\
\mathbf{g}^{\psi\psi} & = \frac{1}{ \sin^2(\theta/2) r^2} + O(r^{-4}), \qquad \mathbf{g}^{\phi\phi} = \frac{1}{\cos^2(\theta/2) r^2} + O(r^{-4}), \qquad \mathbf{g}^{\phi\phi} = O(r^{-6}).
\end{align} Using the expression
\begin{equation}
\star \td m_\psi\vert_S = \left[\partial_r \mathbf{g}_{t\psi} (-\mathbf{g}^{tt}\mathbf{g}^{rr}) + \partial_r \mathbf{g}_{\psi\psi} (-\mathbf{g}^{rr}\mathbf{g}^{\psi t}) + \partial_r \mathbf{g}_{\psi\phi} (-\mathbf{g}^{rr} \mathbf{g}^{t\phi})\right] \td \text{Vol}.
\end{equation} with the $r\to\infty$ limit understood, a  long computation shows that
\begin{equation}
J_\psi^K  = c_1 J_\psi
\end{equation} which fixes $c_1 = 1$. 

A Smarr-type relation can be easily derived using Ricci flatness and the Komar definitions of mass and angular momenta: an application of Stokes' theorem gives
\begin{equation}\begin{aligned}
-\frac{3}{32\pi} \int_S \star \td \xi &= -\frac{3}{32\pi} \int_S \star \td ( k + \Omega_\psi  m_{\psi} + \Omega_\phi m_{\phi}) = \frac{-3}{32\pi}\int_H \star \td \xi  = \frac{3}{16\pi} \kappa_g A_H ,
\end{aligned}
\end{equation} where the volume integral over $\Sigma$ vanishes due to Ricci flatness. This yields 
\begin{equation}
m  = \frac{3}{2} \Omega_\psi J_\psi + \frac{3}{16\pi} \kappa_g A_H .
\end{equation} which is the same identity that holds for stationary, bi-axisymmetric asymptotically flat black holes. It can be verified explicitly. We can also verify by direct computation that the standard first law holds, namely
\begin{equation}
\td m = \frac{\kappa \td A_H}{8\pi} + \Omega_i \td J
\end{equation} where we consider variations in the space of solutions parameterized by $(c,\kappa)$. This requires the use of the regularity conditions fixing $a,b$ in terms of $c$ explicitly.  It is interesting that this variational identity does not include contributions from other geometric features of the solution, such as the disc topology region that lies in the black hole exteriror. 

\subsection{Phase diagrams}
Define the \emph{reduced area} and \emph{reduced angular momentum} 
\begin{align}
    a_H &= \frac{A_H}{m^{3/2}} = \frac{16}{3-c} \sqrt{\frac{2\pi}{3}} \sqrt{(1-c)(3+c)}, \\ 
    j^2 &= \frac{27 \pi J_\psi^2}{32 m^3} = \frac{(3+c)^3}{(3-c)^2(3 + 5c)}
\end{align} Notice that $\kappa$ has been scaled out, so these are dimensionless functions of only $c$. The plot below represents how the area changes with angular momenta while holding the mass fixed.
\begin{figure}[h!]
    \centering
    \includegraphics[width=0.4\linewidth]{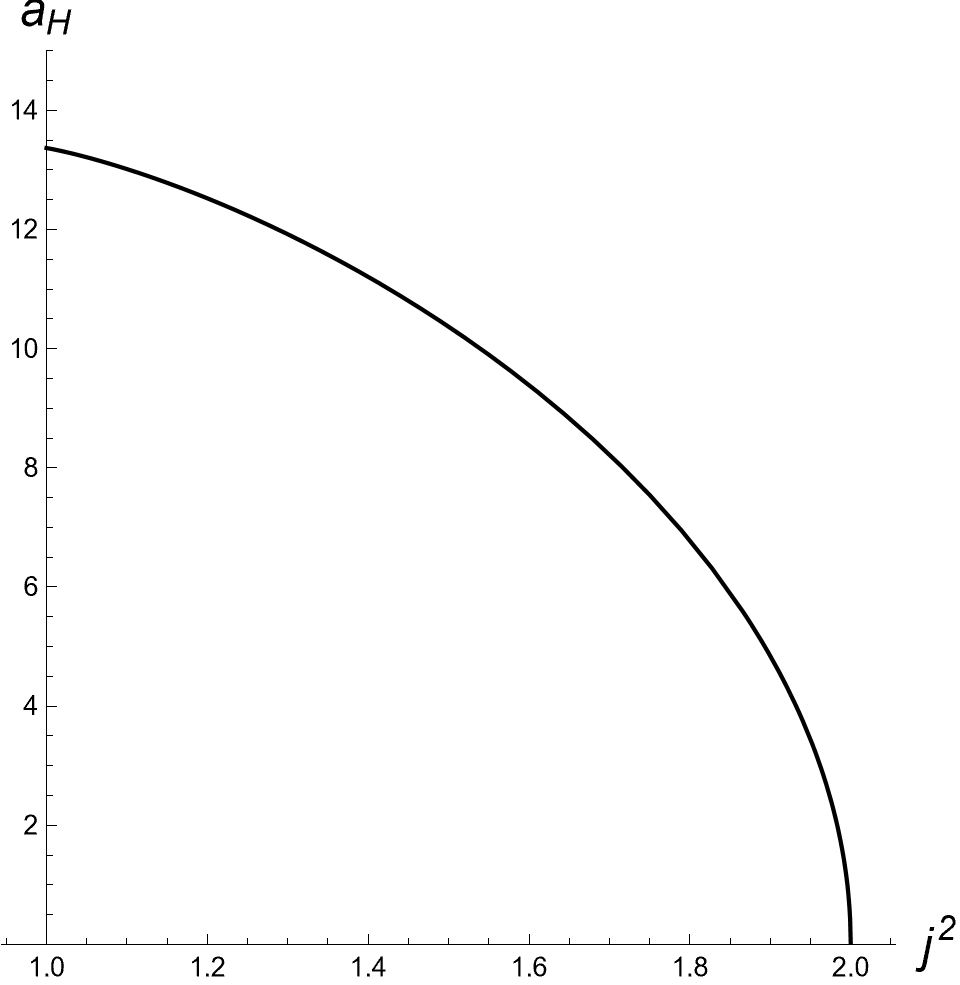}
    \caption{Plot of horizon area vs (spin$)^2$, for given mass, for the rotating Chen-Teo ALE black hole. The maximum reduced area is found at $a_H = \frac{16\sqrt{2\pi}}{3}$ when the reduced angular momentum is at a minimum $j^2 = 1$. As the reduced angular momentum $j^2 \to 2$, $a_H \to 0$ and spacetime contains a naked singularity.}
    \label{fig:enter-label}
\end{figure} It is also useful to understand the behavioiur of the area of the disc (at fixed mass) as the parameter $c$ is varied. 
\begin{figure}[h!]
    \centering
    \includegraphics[width=0.5\linewidth]{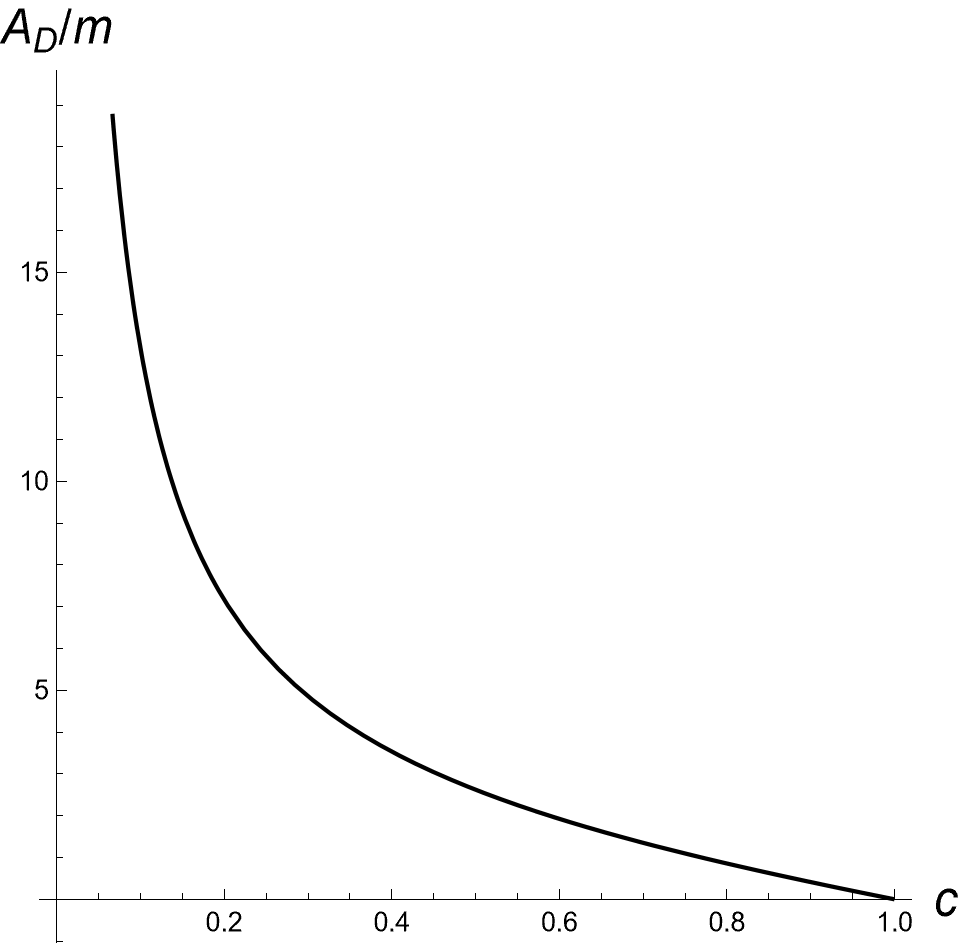}
    \caption{Plot of disc area at fixed mass vs parameter $c$, for the rotating black hole. The dimensionless area diverges as $c \to 0$ and when $c \to 1$ then the disc region vanishes $A_D \to 0$.}
    \label{fig:enter-label2}
\end{figure}

\subsection{Renormalized gravitational action}
We have been unable to find an analytic continuation of the Chen-Teo ALE black hole to a purely Riemannian solution. This is similar to the situation for black ring metrics. However, a (real) gravitational action can still be computed for a complex metric \cite{Brown:1990fk}. A second problem is that the action will be divergent and must be appropriately renormalized by either a background subtraction or counter term method. Here we follow the latter approach following the strategy outlined in \cite{Kraus:1999di} and applied to (complex) dipole black ring metrics in \cite{Astefanesei:2005ad}.
The gravitational action can be computed as a sum of the bulk and \\boundary terms as  
\begin{equation}\label{gravaction}
I_{GH} = \frac{1}{16\pi G}\int_M R\sqrt{-g} \; d^{5}x - \frac{1}{8\pi G} \int_{\partial M} K\sqrt{g} \; d^{4}x
\end{equation}
For these solutions of the instanton, they satisfy the vacuum Einstein solutions namely $R = 0$ so that for extrinsic curvature $K$
\begin{equation}
I_{GH} = - \frac{1}{8\pi G} \int_{\partial M} K\sqrt{g} \; d^{4}x
\end{equation} We consdier a quasi-Euclidean (complex) metric obtained by formally replacing $t \to -i\tau$ in \eqref{ADMLor}. This produces the complex metric
\begin{equation}\label{ADM:complex}
    \mathbf{g}_C = \alpha^2 \td \tau^2 + g_{ij} (\td x^i - i N^i \td \tau)(\td x^j - i N^j \td \tau)
\end{equation} Regularity of the metric requires that the Killing field 
\begin{equation}
    \tilde{\xi} = \frac{1}{\kappa_g} (\partial_\tau + \Omega_\psi \partial_\psi + \Omega_\phi \partial_\phi)
\end{equation} that degenerates on the $S^3$ located at $y = -1/c$, generates $2\pi-$periodic flows. We define the inverse temperature $\beta:= 2\pi / \kappa_g$ (note that the $\tau$ coordinate is not itself periodic, but rather the torus spanned by $(\tau, \psi, \phi)$ satisfies the identification $(\tau, \psi,\phi) \sim (\tau + \beta, \psi + \beta \Omega_\psi, \phi + \beta \Omega_\phi)$ in addition to the identifications $T_1, T_2$ discussed above. It is straightforward to introduce a new angle $\varphi$ adapted to $\tilde\xi$ and then use $(\varphi,\hat\phi^1, \hat\phi^2)$ as a set of independent $2\pi-$periodic angles. 

For Ricci flat metrics, the computation of the gravitational action reduces to calculating a boundary integral over $\partial M$, which is taken to be a level set $r = r_0$, $r_0 \to \infty$, where the asymptotic radial coordinate $r$ defined above.  In particular, one must compute the extrinsic curvature of $\partial M$ embedded in $M$. This is most easily computed by expressing \eqref{ADM:complex} in terms of an ADM decomposition over the radial variable: 
\begin{equation}
\mathbf{g}_C = \tilde{\alpha}^2dr^2 + \tilde{g}_{ij}(dx^i + \tilde{N}^idr)(dx^j + \tilde{N}^jdr)
\end{equation}
Then the extrinsic curvature is given by
\begin{equation}
\tilde{K}_{ij} =  \frac{1}{2\tilde\alpha}(\partial_r \tilde g_{ij} - D_i\tilde N_j - D_j\tilde N_i)
\end{equation} with indices raised and lowered with respect to the auxiliary metric $\tilde{g}$. 
We see that the only nonzero component of the contravariant shift vector is $N_\theta = g_{r\theta}$ which leads to a  simpler expression for $\tilde{K}_{ij}$:
\begin{align*}
\tilde{K}_{ab} = \frac{1}{2\alpha}(\partial_r g_{ab} - N^\theta\partial_\theta g_{ab}) && \tilde{K}_{\theta\theta} = \frac{1}{2\alpha}(\partial_r g_{\theta\theta} - 2g_{\theta\theta}\partial_\theta N^\theta - N^\theta\partial_\theta g_{\theta\theta})
\end{align*} where $(a,b) = \{\tau, \psi,\phi\}$. This leads to following non-zero components
\begin{align*}
\tilde{K}_{\tau\tau} &= \frac{4 b (c-1) \kappa ^2}{(b-1) r^3}+O\left(\frac{1}{r^4}\right)\\
\tilde{K}_{\tau\psi} &= -\frac{4 i \sqrt{2} \left(a^2-1\right) \kappa ^3 \sqrt{\frac{b (b+1) (b-c)}{\left(a^2-1\right) (b-1)}} (1-c) \sin^2\left(\frac{\theta}{2}\right)}{(b-1) r^3}+O\left(\frac{1}{r^4}\right), \quad 
\tilde{K}_{\tau\phi} = O\left(\frac{1}{r^7}\right)\\
\tilde{K}_{\psi\phi} &= \frac{8 a \kappa ^4 \left(a^2 b+a^2+b-1\right) (1-c) \left(a^2 b-a^2 c-b c+c\right) \sin^4 \left(\frac{\theta}{2}\right)}{r^3 \left(\left(a^2-1\right)^2 (b-1)^2\right)} +O\left(\frac{1}{r^4}\right)\\
   \tilde{K}_{\psi\psi} &=  r \sin^2\left(\frac{\theta}{2}\right) +O\left(1\right),\quad
   \tilde{K}_{\phi\phi} = r \cos^2 \left(\frac{\theta}{2}\right) +O\left(1\right),\quad
   \tilde{K}_{\theta\theta} = \frac{r}{4} +O\left(1\right).
\end{align*}
Then the mean curvature (trace of the extrinsic curvature) and the volume element of the hypersurface are respectively
\begin{equation}
\tilde{K} = \frac{3}{r} - \frac{5\kappa^2 b (1 - c)}{(1 - b)r^3} = \frac{3}{r} - \frac{20}{3r^3\pi}m + O\left(\frac{1}{r^5}\right) 
\end{equation}
\begin{equation}
\sqrt{\det \tilde{g}} = \frac{r^3}{4}\sin \theta  + \frac{r}{3\pi} m\sin \theta + O\left(1\right)
\end{equation}
Then the divergent action is calculated to be the $r \to \infty$ limit of 
\begin{equation}
I_{GH} = \frac{1}{8 \pi}\int_{\partial M}\tilde{K}\sqrt{\det{\tilde{g}}}\; \td^4x  = -\frac{3r^2\pi\beta}{8} + \frac{m}{3}\beta.
\end{equation} Here, the integration over the torus is taken over $\td \tau \wedge \td \psi \wedge \td \psi = (\beta / 4\pi) \td \varphi \wedge \td \hat \phi^1 \wedge \td \hat \phi^2$. 
Note that this is divergent as $r \rightarrow \infty$. Therefore we must implement an appropriate renormalization procedure. We will follow the counterterm method outlined in \cite{Kraus:1999di} for spaces that are asymptotic to $S^1 \times S^3$. The only modification is that the $S^3$ must be replaced by $L(2,1)$ in our setting (note also that the $S^1$ has infinite size because the orbits of $\partial_\tau$ do not close, but the volume growth is still quartic and so the integrals converge). To proceed we first calculate the intrinsic Ricci curvature for this metric on the boundary surface as 
\begin{equation}
R = \frac{6}{r^2} - \frac{16}{\pi r^4}m + O\left(\frac{1}{r^5}\right) \implies \sqrt{R} = \frac{\sqrt{6}}{r} - \frac{4\sqrt{6}}{3\pi r^3}m + O\left(\frac{1}{r^4}\right)
\end{equation}   The renormalized gravitational action $I_{\text{ren}}$ is obtained by supplementing \eqref{gravaction} with an additional term proportional to the Ricci scalar of the boundary metric induced on $\partial M$: 
\begin{equation}
I_{\text{ren}} := \frac{1}{8\pi}\int_{\partial M}  \left(\tilde{K} - \sqrt{\frac{3}{2}}\sqrt{R}\right)\; \sqrt{\tilde g}\; \td^4x = \frac{1}{8\pi}\int_{\partial M}  \left(-\frac{20}{3} + 4\right)\frac{m}{\pi r^3}\; \sqrt{\tilde g}\; \td^4x = \frac{m}{3}\beta
\end{equation} Using the equation for the Gibbs potential, and the Smarr relation, we obtain the standard thermodynamic identity 
\begin{equation}
G(T, \Omega_\psi) = \beta^{-1} S = \frac{m}{3} = m - \frac{2}{3}m = m - \Omega_\psi J_\psi - TS
\end{equation} where $T = \beta^{-1}  = \kappa / 2\pi$ is the Hawking temperature and $S = A_H/4$ is the entropy of the black hole.  From here we can read off the extensive thermodynamic variables (entropy and angular momentum) from the standard expressions
\begin{equation}
    S = -\frac{\partial G}{\partial T}\Bigg|_\Omega, \qquad J = - \frac{\partial G}{\partial \Omega}\Bigg|_{T}.
\end{equation} The positivity of the free energy, represented by the gravitational action, indicates that the thermodynamically preferred state (with minimum action) is the solution with $m =0$. The natural candidate for the dominant contribution to the partition function is not the orbifold $S^1 \times \mathbb{R}^4 / \mathbb{Z}_2$ (which has an orbifold singularity at the origin) but rather a thermal ensemble on the smooth spatial Eguchi-Hanson space $M_{EH}$. The latter can be described by a path integral on the Riemannian Ricci flat metric on the Eguchi-Hanson `soliton' $S^1 \times M_{EH}$ where the $S^1$ has radius $\beta$. It will be interesting to investigate whether this conclusion will hold for other black holes on different gravitational instanton backgrounds. 
\section{Discussion}
The horizon topology theorem of Galloway-Schoen proves that spatial cross-sections $H$ of the event horizon admit metric with positive scalar curvature \cite{Galloway:2005mf}. In five dimensions, the only possibilities are spherical spaces (three-manifolds covered by $S^3$, includes lens spaces), $S^1 \times S^2$, and connected sums thereof. In an asymptotically flat stationary black hole spacetime, one can think of a spatial hypersurface in the domain of outer communications as a four manifold $\Sigma$ with inner boundary $H$ and asymptotic boundary $S^3$ (a cobordism). It can be in fact be proved that $\Sigma$ is simply connected \cite{Eichmair:2012jk}. This is a consequence of the topological censorship theorem which constraints the fundamental group of the domain of outer communications in the terms of the topology of a neighbourhood of null infinity \cite{Friedman:1993ty} (the result can be suitably extended into the setting of initial data alone \cite{Andersson:2015sfa, Eichmair:2012jk}). For example, in a black ring spacetime, $\Sigma$ represents a cobordism between $H = S^1 \times S^2$ and $\partial_\infty\Sigma = S^3$ \cite{Alaee:2013oja}.  For a black lens spacetime \cite{Kunduri:2014kja,Tomizawa:2016kjh} , the corresponding $\Sigma$ is a cobordism between $L(n,1)$ and $S^3$  The Chen-Teo ALE solution studied in this work can then be loosely thought of as a `reverse' of this, in the sense that now $H = S^3$ and the asymptotic boundary is $L(2,1)$. It would be an interesting problem to see what constraints are placed on the topology of the domain of outer communications when the asymptotic end is that of a gravitational instanton (in particular, the end need not be simply connected). 

For asymptotically flat spherical biaxisymmetric black hole (maximal) initial data sets the following inequality holds \cite{Alaee:2015pwa}:
\begin{equation}\label{MJ}
m^3 \geq \frac{27\pi}{32} (|J_1| + |J_1|)^2
\end{equation} where the $J_i$ correspond to angular momenta associated to independent $2\pi-$periodic rotations in two orthogonal planes of rotation at spatial infinity (see \cite{Alaee:2017ygv} for an analogous geometric inequality satisfied by black ring initial data).  Equality is achieved if and only if the initial data is that of a canonical time slice of (the unique) extreme Myers-Perry black hole with this $(J_1, J_2)$.  The ALE black hole solutions studied here must have angular momentum along one orthogonal direction, while in another orthogonal plane, the angular momentum vanishes, and there is no extreme limit. It would be interesting to use the inverse scattering methods to find a more general family of solutions with two independent angular momenta; such a family would presumably contain at least one branch of degenerate subfamilies with vanishing surface gravity. These extreme geometries could be minimizers for geometric inequalities satisfied by biaxisymmetric initial data modeling ALE black holes. Observe that using the transformations \eqref{anglechange}, we can read off $J_{\hat\phi^1} = J_\psi/2$, $J_{\hat\phi^2} = J_\phi =0$ where $(\hat\phi^1, \hat\phi^2)$ are the generators of the torus action with independent $2\pi-$periodic flows. In that case, it can be checked that the above (strict) inequality \eqref{MJ} is satisfied with $(J_1,J_2)$ identified with $(J_{\hat\phi^1}, J_{\hat\phi^2})$ respectively.  However, $(\partial_{\hat\phi^1}, \partial_{\hat\phi^2})$ are not orthogonal at spatial infinity, so the above identification may not be natural. 

We have studied the thermodynamic properties of the Chen-Teo family of ALE black holes and shown they satisfy the expected Smarr and first law black hole mechanics formulae. Since these are stationary, biaxisymmetric solutions, it should be possible to derive refined divergence-type identities for ALE black holes as was done in the asymptotically flat setting \cite{Kunduri:2018qqt}; this problem is under current investigation. We have also shown that these black holes have positive free energy relative to a thermal Eguchi-Hanson soliton background, suggesting a thermal instability analogous to the well known Schwarzschild example. It would be interesting to see how this is connected to dynamical instabilities.  A simple toy model to investigate this would be to determine the behaviour of a scalar field in this background.

\bibliographystyle{unsrt}
\bibliography{refs} 
\end{document}